\documentclass[aps,pra]{revtex4-1}
\usepackage{natbib}
\usepackage{amsmath,amsfonts,amssymb}%
\usepackage{graphicx,color}

\usepackage{float}
\usepackage[colorlinks,linkcolor=black,anchorcolor=black,citecolor=black]{hyperref}
\draft 

\begin{document}

\title{Scattering of flexural waves from an $N$-beam resonator in a thin plate}



\author{Alfonso Climente$^1$}
\author{Penglin Gao$^{1,2}$}
\author{Linzhi Wu$^2$}
\author{Jos\'e S\'anchez-Dehesa$^1$}
\email[Corresponding author: ]{jsdehesa@upv.es}
\affiliation{$^a$Wave Phenomena Group, Department of Electronic Engineering, Universitat Polit\`ecnica de Val\`encia, Camino de vera s.n. (Building 7F), ES-46022 Valencia, Spain.\\
$^b$Center for Composite Materials, Harbin Institute of Technology, Harbin 150001, China.}

\date{\today}

\begin{abstract}
The impedance matrix method is applied to study the scattering of flexural waves propagating in an infinite thin plate containing an $N$-beam resonator.
The resonator consists of a circular hole containing a smaller plate connected to the background plate by a number $N$ of rectangular beams.
After representing the boundary conditions in a modal multipole expansion form, a compact expression is obtained for the T-matrix, which relates the incident and the scattered transverse (out-of-plane) waves.
The analysis of the scattering cross-section reveals interesting scattering features, like resonances and anisotropy, associated to this type of resonators.
Numerical experiments performed within the framework of the finite element method support the accuracy of the model here developed.
\end{abstract}


\maketitle 

\section{Introduction}
The propagation of waves in periodically structured plates is currently a topic of increasing interest due to their potential applications for vibration control.
The reader is addressed to recent reviews\cite{hussein2014dynamics,wu2011phononic} and references therein for a comprehensive report on this topic.
 When the plate is thin enough, the wavelength is much larger than the plate's thickness and, consequently, the lower order transversely polarized plate wave modes dominate.
Particularly, the first order asymmetric Lamb mode is termed as flexural wave, and it is governed by a biharmonic equation based on Kirchhoff-Love plate theory.\citep{graff2012wave}
One advantage of these particular waves is that the deformation of a plate can be easily obtained by optical measurement systems, which facilitates the experimental validation of the  devices based on them.
Many works have been reported for flexural waves concerning band gaps,\citep{mcphedran2009platonic,hsu2006efficient,vasseur2008absolute} negative refraction and focusing,\citep{farhat2010focussing,wu2011focusing,bramhavar2011negative} cloaking,\citep{farhat2012broadband,stenger2012experiments,climente2016analysis} gradient index lenses \cite{climente2014gradient,jin2016gradient} and homogenization.\citep{torrent2014effective}
\par
The bandgaps obtained from the Bragg scattering by periodically arranged scatterers in plates appear in frequency region where the wavelength is comparable to the lattice constant.
For this reason, they can hardly be used in practice because of the unfeasible dimensions of the samples to be manufactured if one wants to get low frequency bandgaps.
To overcome this difficulty, flexural bandgaps at extremely low frequencies can be obtained by designing periodic structures with embedded resonant structures, imitating the seminal idea introduced by Liu et \emph{al.} \citep{liu2000locally} in acoustics.
The resulting periodic systems, composed of local resonators, are sometimes named as platonic \citep{mcphedran2009platonic} or elastic metamaterials. \citep{zhu2015microstructural}
 In the range of working frequencies, the wavelength is much larger than the lattice constant, hence metamaterials are often characterized by frequency related effective parameters just like they were homogeneous.\citep{zhu2014negative,torrent2014effective}
Subsequently, various kinds of negative index metamaterials were proposed and designed by means of local resonance mechanism, which paves the way to some intriguing phenomena, like negative refraction and perfect lensing. \citep{zhu2015microstructural}
\par
For flexural waves, the local resonances are usually produced by spring-mass like resonators vibrating in transverse (out-of-plane) direction, such as spring-masses,\citep{xiao2011formation} pillars,\citep{wu2008evidence,pennec2008low,oudich2011experimental,khelif2010locally} and holes with inner structures.\citep{andreassen2015directional,farhat2014biharmonic}
Among these works, finite element method was widely used on account of the complexity of resonant structures, and only few works concerning spring-mass resonators were carried out by analytical methods like the plane wave expansion method.\citep{xiao2011formation}
For a comprehensive report, the reader is addressed to the review article by Zhu and coworkers.\citep{zhu2015microstructural}
\par
This work presents an analytical formulation for the scattering of flexural waves from a single structured circular hole consisting of a circular plate that is connected to the background plate with $N$ rectangular beams.
These type of resonators, which are named as $N$-beam resonators, have been already employed to predict directional wave motion \cite{andreassen2015directional} and superlensing. \cite{farhat2014biharmonic}
There are two main reasons justifying the study of these resonators.
On one hand, they are of easy manufacturing by using laser or waterjet cutting machines.
On the other hand, the resonators can be modeled analytically by coupling the Kirchhoff-Love and the Euler-Bernoulli theories as it is shown below.
Particularly, the model here developed is an extension of the impedance matrix method previously used in solving the scattering of flexural waves from a hole with an internal beam. \citep{climente2015scattering}
In this paper, a similar procedure is followed to get the transfer matrix (T-matrix) of the complex resonator under study, which lays a foundation on analysing metamaterials with multiple scattering theory.

The paper is organized as follows. After this introduction, Sec. \uppercase\expandafter{\romannumeral2} gives the detailed procedures for obtaining the T-matrix with impedance matrix method.
In Sec. \uppercase\expandafter{\romannumeral3}, several numerical simulations are shown to validate the present theory and to illuminate some complex scattering characteristics, like resonance and anisotropy. Finally, the work is summarized in Sec. \uppercase\expandafter{\romannumeral4}, which also gives an outlook of potential applications of the model here introduced.
\section{Scattering of flexural waves from a $N$-beam resonator}
\subsection{Problem definition}
\begin{figure}
    \centering
    \includegraphics[width=\linewidth]{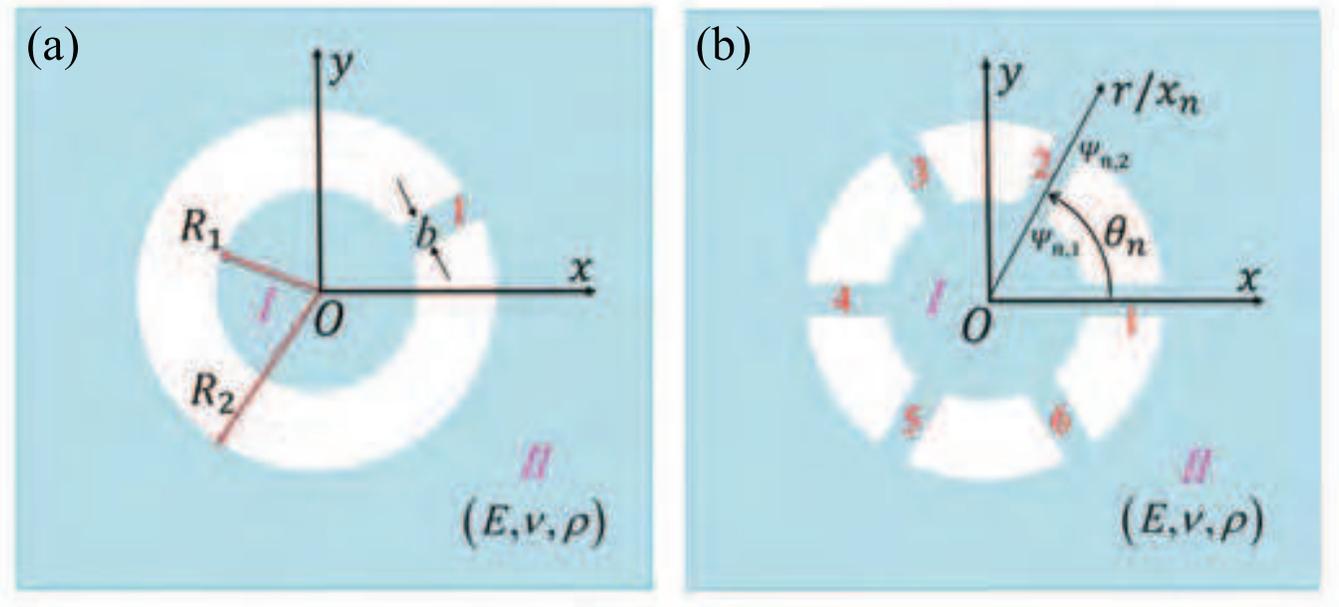}
	\caption{(color on line) (a) Schematic diagram of a hole resonator consisting of a circular disk (region I) connected to the background plate (region II) by a single rectangular beam.
	(b) The case of a resonator connected with six beams, where $\theta_n$ defines the inclination angle of the $n$-th beam.
	In both cases, the radii of hole and disk are $R_2$ and $R_1$, respectively. The width of all the beams is $b$ and the anchor points at the $n$-th beam are $\Psi_{n,1}$ and $\Psi_{n,2}$.}
    \label{fig1}
\end{figure}

Figure \ref{fig1} shows two examples of the $N$-beam resonators under study.
The case $N=$1, the single beam resonator, is depicted in Fig. \ref{fig1}(a), which shows a hole containing a disk or inner plate (region I) connected to the background plate (region II) by just one beam.
Figure \ref{fig1}(b) corresponds to $N=$6, representing the case with the maximum number of beams for which numerical results are reported here.
The beams, which are rectangular with length $R_2-R_1$ and width $b$, are uniformly distributed inside the hole.
However, let us remark that the developed theory applies to a general case with arbitrary angles.
Anchor points are named for the $n$-th beam as $\Psi_{n,1}$ for region I and $\Psi_{n,2}$ for region II.
For convenience, the origins of the Cartesian coordinates $O$-$xy$ and polar coordinates $O$-$r\theta$ are placed at the center of region I.
Each beam is described in a local system $Ox_{n}$, which is defined along the $n$-th beam's axis and pointing the $r$-direction, as shown in Figure \ref{fig1}(b).
It is assumed that the resonant scatterer is fabricated by a waterjet cutting machine and all parts are made of the same material as the background plate with density $\rho$ and rigidity $D=Eh^3/12\left(1-\nu^2\right)$, where $E$ and $\nu$ are the Young's modulus and Poisson's ratio, respectively. Moreover, $h$ is the thickness of the plate.
\par
According to plate theory, if $h$ is thin enough the flexural (transverse) waves, $W$, can be obtained by solving Kirchhoff-Love biharmonic equation\citep{graff2012wave}
\begin{equation}
\nabla^4W-k_p^4W=0,
\end{equation}
\noindent where $k_p=(\rho h\omega^2/D)^{1/4}$ is the wave number with $\omega$ being the angular frequency.
The time harmonic factor $e^{-i\omega t}$ is implicit in all the formulation but will be omitted throughout the paper for simplicity.
\par
In polar coordinates, the solutions for the inner plate ($W^{\uppercase\expandafter{\romannumeral1}}$) and background plate ($W^{\uppercase\expandafter{\romannumeral2}}$) can be expressed as
\begin{align}
        \label{W1}
		W^{I}(r,\theta)&=\sum_{q=-\infty}^{\infty} \left [ C^{(J)}_q J_q(k_p r) + C^{(I)}_q I_q(k_p r)\right ] e^{i q\theta},\\
        \label{W2}
        W^{II}(r,\theta)&= \sum_{q=-\infty}^{\infty} \left [ A^{(J)}_q J_q(k_p r) + A^{(I)}_q I_q(k_p r)+B^{(H)}_q H_q(k_p r) + B^{(K)}_q K_q(k_p r)\right ] e^{i q\theta},
\end{align}
\noindent where $J_{q}(\cdot)$ is the Bessel function, $H_{q}(\cdot)$ is the Hankel function of first kind, and $I_{q}(\cdot)$, $K_{q}(\cdot)$ are the modified Bessel functions.
$A_{q}^{(*)}$, $B_{q}^{(*)}$ and $C_{q}^{(*)}$ represent the expansion coefficients for incoming wave, scattered wave and internal wave, respectively.
For the numerical simulations, the summations run from $-N_q$ to $N_q$, the value $N_q$ giving the adequate truncation of the series is determined for each resonator and mainly depends on the resonator dimension and the wavelength (see Sec. \ref{convergence}).
\par
\subsection{Wave solution within the beams and boundary conditions}
The transverse (out-of-plane) displacement inside the $n$-th beam, $V^{n}(x)$, can be obtained by solving the corresponding equation of motion
\begin{equation}
        \frac{\partial^4 V^{n}}{\partial x^4}-k_b^4V^n =0,
\end{equation}
where the beam is defined for $x\in[R_1,R_2]$ with an inclination angle of $\theta _n$.
The wave number in the beam is $k_b=(m\omega^2/EI)^{1/4}$, where $I=bh^3/12$ is the second moment of area and $m=\rho bh$ is the mass per unit length.
The solution can be expressed as
\begin{equation}
\label{eq_beam}
V^{n}(x) = D_1^n  e^{ik_b x}+D_2^n  e^{-ik_b x}+D_3^n  e^{k_b x}+D_4^n e^{-k_b x},
\end{equation}
\noindent where coefficients $D_i^n$ ($i=$ 1 to 4) define the contributions of the different waves inside the beams.
\par
The interaction between waves propagating in the beams with those propagating within plates I and II is embodied through the boundary conditions.
It is clear that the radial moment $M_r$ and radial Kirchhoff stress $V_r$ are vanishing for a free hole, but for the problem here considered, some additional conditions should be introduced at the anchor points.
A detailed account of the basic formulation describing the connection between beam and plate can be found in Ref. \onlinecite{climente2015scattering}.
Obviously, the displacement, the slope, the moment and shear stress, all of them should be continuous at both anchor points for the $n$-th beam.
 The mathematical expressions for these boundary conditions are
\begin{subequations}
\begin{equation}
		\label{bda}
    W^{I}(R_1,\theta_{n}) = V^{n}(R_1)   \quad \text{and} \quad  W^{II}(R_2,\theta_{n}) = V^{n}(R_2),
\end{equation}
\begin{equation}
    \label{bdb}
	\left . \frac{\partial W^{I}}     {\partial r} \right |_{\substack{r= R_1 \\ \theta=\theta_{n}} } =
	\left . \frac{\partial V^{n}} 		{\partial x} \right |_{x= R_1} \quad
	\text{and} \quad
	\left . \frac{\partial W^{II}}    {\partial r} \right |_{\substack{r= R_2 \\ \theta=\theta_{n}} } =
	\left . \frac{\partial V^{n}}			{\partial x} \right |_{x= R_2},
\end{equation}

\begin{equation}
    \label{bdc}
	\left . 			M^{I}_r 		       \right |_{\substack{r= R_1 \\ \theta=\theta_{n}}  }=
	\left . \frac{M^{n}_x}{R_1}  \right |_{x= R_1} \quad
	\text{and} \quad
	\left . 			M^{II}_r 	         \right |_{\substack{r= R_2 \\ \theta=\theta_{n}} }=
	\left . \frac{M^{n}_x}{R_2} \right |_{x= R_2},
\end{equation}
\begin{equation}
    \label{bdd}
	\left . 			V^{I}_r 		       \right |_{\substack{r= R_1 \\ \theta=\theta_{n}}  }=
	\left . \frac{Q^{n}_x}{R_1}  \right |_{x= R_1} \quad
	\text{and}  \quad
	\left . 			V^{II}_r 		       \right |_{\substack{r= R_2 \\ \theta=\theta_{n}} }=
	\left . \frac{Q^{n}_x}{R_2} \right |_{x= R_2}.
\end{equation}
\end{subequations}
In these formulas, $M_x^{n}$ and $Q_x^{n}$ are the bending moment and the transverse shear force, respectively.
\subsection{T-matrix solution}\label{Tmatrix}
The transfer matrix or T-matrix is the key ingredient in order to solve any scattering problem and it allows obtaining the scattered field for any type of incident wave.
The T-matrix relates the incoming and scattered wave coefficients through the relationship
\begin{equation}
\label{Tmatrix}
B=\boldsymbol{T}A,
\end{equation}
\noindent where $A$ and $B$ are column vectors containing the incoming and scattered wave coefficients of Eq. (\ref{W2}).
Their elements are $A_q=[A_q^{(J)} A_q^{(I)}]^t$ and $B_q=[B_q^{(H)}B_q^{(K)}]^t$, respectively.
Therefore, the $\boldsymbol{T}$ is a square matrix that for numerical calculations has dimension $2(2N_q+1)\times 2(2N_q+1)$.
\par
By applying the impedance method for flexural waves, \citep{climente2015scattering} the T-matrix of the $N$-beam resonator can be cast in:
\begin{equation}
    \label{T-matrix}
    \boldsymbol{T} =
    [ \boldsymbol{M}^{HK}(k_p,R_2) ]^{-1}\
    [\boldsymbol{Z}^{scat}+\boldsymbol{Z}^{tot}]^{-1}\
    [\boldsymbol{Z}^{inc}-\boldsymbol{Z}^{tot}]\
    [\boldsymbol{M}^{JI}(k_p,R_2)],
\end{equation}
\noindent where the matrices $\boldsymbol{Z}^{inc}$, $\boldsymbol{Z}^{scat}$ and $\boldsymbol{Z}^{tot}$ are evaluated at the boundary $r=R_2$.
They represent the impedance matrices associated to the incident ($\boldsymbol{Z}^{inc}$), scattered ($\boldsymbol{Z}^{scat}$) and total ($\boldsymbol{Z}^{tot}$) flexural waves propagating in region II.
Their expressions are given in Appendix \ref{plate} and  \ref{plate2}, which also contain an explanation of their derivation.
\par
Once the $\boldsymbol{T}$ is determined, coefficients $B^{(*)}_q$ are obtained from,
\begin{equation}
    \begin{bmatrix} B_q^{(H)} \\ B_q^{(K)} \end{bmatrix}=\sum_{s=-\infty}^{\infty}
    \boldsymbol{T}_{qs}\begin{bmatrix} A_s^{(J)} \\ A_s^{(I)} \end{bmatrix},
\end{equation}
which fully determines $W^{II}$, the waves propagating in region II, since coefficients $A_q^{(*)}$ defining the incident wave are known.
For the cases considered here, an incident plane wave and a punctual source, the expressions for $A_q^{(*)}$ can be found elsewhere. \cite{climente2015scattering}
\par
Now, the coefficients $C_q^{(*)}$, determining the flexural waves propagating in the inner plate, $W^I$, are obtained with further mathematical operations.
The combination of Eqs. (\ref{bda}), (\ref{bdb}), (\ref{stiffness}), (\ref{innerplateW}), (\ref{innerplateM}) and (\ref{Yqq2}) yields
\begin{equation}
    \sum_{s=-\infty}^{\infty} \boldsymbol{\hat{L}}_{qs}
    \begin{bmatrix} C_s^{(J)} \\ C_s^{(I)} \end{bmatrix}=\sum_{s=-\infty}^{\infty} \boldsymbol{\hat{R}}_{qs}\begin{bmatrix} W^{\uppercase\expandafter{\romannumeral2}} \\ W_{,r}^{\uppercase\expandafter{\romannumeral2}} \end{bmatrix}_{s},
\end{equation}
with $W^{II}_{,r}\equiv\partial W^{II}/\partial r$ and where the $2\times 2$ sub-matrices $\boldsymbol{\hat{L}_{qs}}$ and $\boldsymbol{\hat{R}_{qs}}$ are the elements of $\boldsymbol{\hat{L}}$ and $\boldsymbol{\hat{R}}$, respectively.
\begin{equation}
    \boldsymbol{\hat{L}}_{qs}=2\pi R_1 \boldsymbol{N}_{ss}^{JI}(k_p,R_1) \boldsymbol{\delta}_{qs}-
    \boldsymbol{K}_{11} \boldsymbol{M}_{ss}^{JI}(k_p,R_1) \sum_{m=1}^{N}e^{i(s-q)\theta_m},
\end{equation}
\begin{equation}
    \boldsymbol{\hat{R}}_{qs}=\boldsymbol{K}_{12} \sum_{m=1}^{N} e^{i(s-q)\theta_m}.
\end{equation}
\par
Finally, by solving the above equation we obtain the coefficients in region I as
\begin{equation}
    \begin{bmatrix} C_q^{(J)} \\ C_q^{(I)} \end{bmatrix}=\sum_{s=-\infty}^{\infty}
    \boldsymbol{\hat{T}}_{qs}\begin{bmatrix} W^{\uppercase\expandafter{\romannumeral2}} \\ W_{,r}^{\uppercase\expandafter{\romannumeral2}} \end{bmatrix}_{s},
\end{equation}
where the sub-matrix $\boldsymbol{\hat{T}}_{qs}$ is one element of the block matrix $\boldsymbol{\hat{T}}=\boldsymbol{\hat{L}}^{-1}\boldsymbol{\hat{R}}$.
\par
Once the displacements in the background plate and inner plate are known, according to Eqs. (\ref{bda}) and (\ref{bdb}) the coefficients determining the flexural waves in the $n$-beam are easily available
\begin{multline}
    \begin{bmatrix} D_1^n \\ D_2^n \\ D_3^n \\ D_4^n \end{bmatrix}=\boldsymbol{H}^{-1} \sum_{q=-\infty}^{\infty}
    \begin{bmatrix} \boldsymbol{0} & \boldsymbol{0} \\ \boldsymbol{M}_{qq}^{JI}(k_p,R_2) & \boldsymbol{M}_{qq}^{HK}(k_p,R_2) \end{bmatrix} \begin{bmatrix} A_q^{(J)} \\ A_q^{(I)} \\ B_q^{(H)} \\ B_q^{(K)} \end{bmatrix} e^{iq\theta_n}
    \\
    +\boldsymbol{H}^{-1}\sum_{q=-\infty}^{\infty} \begin{bmatrix} \boldsymbol{M}_{qq}^{JI}(k_p,R_1) & \boldsymbol{0} \\ \boldsymbol{0} & \boldsymbol{0} \end{bmatrix} \begin{bmatrix} C_q^{(J)} \\ C_q^{(I)} \\ 0 \\ 0 \end{bmatrix} e^{iq\theta_n}.
\end{multline}
The expressions of these coefficients totally solve the scattering problem proposed in Eqs. (\ref{W1}), (\ref{W2}) and (\ref{eq_beam}).
\section{Results and discussion}
This section reports numerical simulations checking the convergence and accuracy of the model introduced previously.
In this section, two physical magnitudes are studied.
The first one is the angle dependent {\it scattering amplitude at the far field} \citep{norris1995scattering}
\begin{equation}\label{scatampl}
    f(\theta)=\frac{2}{\sqrt{\pi k_p}}\sum_{q=-\infty}^{\infty}(-i)^{q} B_{q}^{(H)}e^{iq\theta}.
\end{equation}
Particularly, when $\theta=0$, it is called as the {\it forward scattering amplitude} and its modulus $|f(0)|$ is used to perform the study of convergence.
\par
The second magnitude of interest is the scattering cross-section $\sigma_{sc}$, which is defined as the ratio of scattered flux to the incident flux.
It is a useful quantity for analyzing the scattering properties at the far field.
According to the optical theorem for flexural waves, $\sigma_{sc}$ only depends on the scattering coefficients $B_{q}^{(H)}$, \citep{norris1995scattering}
\begin{equation}
    \label{eqcs}
    \sigma_{sc}=-\frac{4}{k_p}\sum_{q=-\infty}^{\infty} {\rm Re}\left[ (-i)^{q}B_{q}^{(H)}\right].
\end{equation}
\par
Numerical results are obtained by considering an infinite plate made of aluminum (Young's modulus $E=\text{69}\,\text{GPa}$, Poisson's ratio $\nu=\,\text{0.33}$, and the density $\rho=\text{2.7}\times\text{10}^3 \, \text{kg} / \text{m}^3 $).
Other parameters described in Fig. \ref{fig1} are: $R_1=\text{5}\,\text{mm}$, $R_2=$ 10 mm, $b=$ 2 mm and the plate thickness (not shown) is $h=$ 1 mm.
These parameters have been taken from Ref. \onlinecite{andreassen2015directional} studying a square lattice of 2-beam resonators and are employed here for comparison purposes.
\subsection{Convergence analysis}\label{convergence}
\begin{figure}[h]
    \centering
    \includegraphics[scale=1]{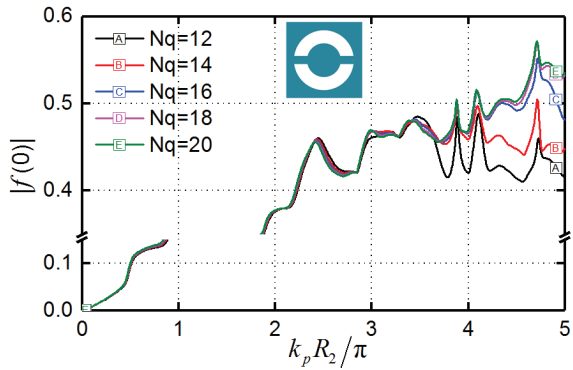}
	\caption{(color on line) Convergence study for the case of a 2-beam resonator.
	The modulus of the forward scattering amplitude, $|f(0)|$,  is plotted as a function of the normalized frequency, $k_pR_2/\pi$, for five different values of $N_q$, the integer defining the series truncation.}
    \label{fig2}
\end{figure}
\par
Figure \ref{fig2} shows the study of convergence for the case of a 2-beam resonator.
The modulus of the forward scattering amplitude at the far field, $f|(0)|$, is calculated as a function of the frequency (in normalized units) for five different values of $N_q$, the integer number defining the truncation of the series expansion in Eq. (\ref{scatampl}).
As expected, it is observed that, for larger frequencies, the convergence is achieved with more summation terms.
Therefore, from Fig. \ref{fig2} it is concluded that for $N_q=12$ the series summation converges fairly well if the normalized frequency is smaller than 3; that is, when the size of the resonator is smaller than $1.5\lambda$, where $\lambda$ is the wavelength inside the uniform plate.
For higher frequencies, a better accuracy is reached at the cost of larger computing time.
A compromise has to be made between accuracy and computing time.
So, in the frequency range explored in this work (up to 5 reduced units), $N_q=18$ is a good choice and will be used in the following calculations.
\subsection{Finite element simulations}\label{sec:fem}
A commercial finite element package (COMSOL Multiphysics) is employed here to study numerically the behavior of two different $N$-beam resonators.
The results are compared with those obtained with the model here developed.
Using the solid mechanics module, the full elastic equations in three-dimensions (3D) are solved, so that both the in-plane and out-of-plane motions are involved in the displacement field.
Under the action of a harmonic excitation perpendicularly applied to the thin plate, the transverse (out-of-plane) component of the displacement dominates and it can be regarded as the flexural wave motion in our model.
In the simulations, the punctual source is located at (-2, 2), where the Cartesian coordinates are normalized to $R_2$.
\begin{figure}[h]
    \centering
    \includegraphics[scale=1]{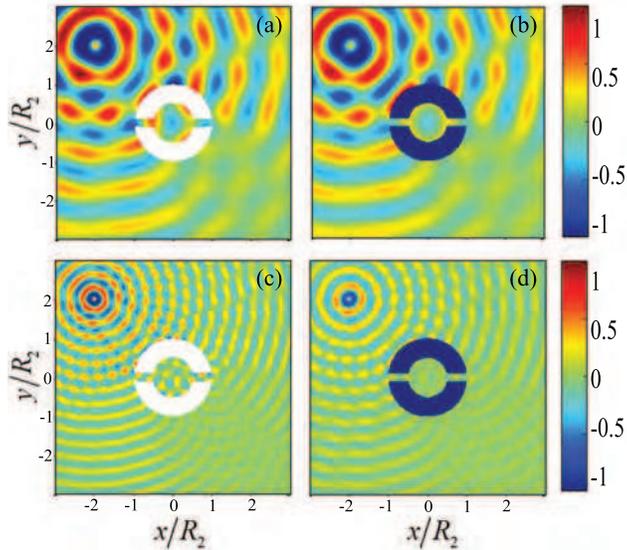}
	\caption{(color on line) Snapshots of the out-of-plane displacements obtained when a cylindrical wave is scattered by a 2-beam resonator.
	The punctual source is placed at ($-$2, 2) and the excitation wavenumbers are $k_{p}R_2=2\pi$ and $k_{p}R_2=5\pi$ for the upper and lower panels, respectively.
	Left panels correspond to 3D finite element simulations and right panels correspond to the analytical 2D model here introduced.}
    \label{fig3}
\end{figure}
\par
Figure \ref{fig3} shows snapshots of the out-of-plane displacements obtained from the {\it exact} 3D finite element simulator (left panels) and the analytical simulations (right panels) when a 2-beam resonator is placed at the plate center.
Results in Figs. \ref{fig3}(a) and \ref{fig3}(b) are obtained by exciting the system with a punctual source with a wavenumber such that $k_{p}R_2=2\pi$.
Their comparison indicates that the displacement patterns are practically the same, showing that the scattering pattern obtained by the 3D solver is perfectly reproduced by our analytical 2D model.
However, the situation is slightly  different for Figs. \ref{fig3}(c) and \ref{fig3}(d), which are calculated with an excitation having $k_{p}R_2=5\pi$.
For this case, even though they show similar scattering patterns, appreciable differences are observed.
The origin of the discrepancy is attributed to the implicit limitations in the Kirchhoff-Love theory employed in our modeling.
When $k_{p}R_2=5\pi$, the wavelength is only four times the thickness of the plate and, therefore, the plate is not thin enough to be treated as a Kirchhoff plate.
\par
\begin{figure}[h]
    \centering
    \includegraphics[scale=1]{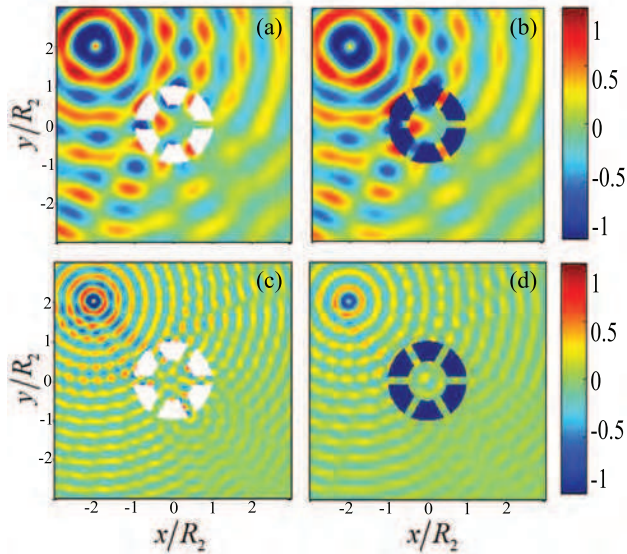}
	\caption{(color on line) Snapshots of the out-of-plane displacements obtained when a cylindrical wave is scattered by a 6-beam resonator.
	The punctual source is placed at ($-$2, 2) and the excitation wavenumbers are $k_{p}R_2=2\pi$ and $k_{p}R_2=5\pi$ for the upper and lower panels, respectively.
	Left panels correspond to 3D finite element simulations and right panels correspond to the analytical 2D model here introduced.}
    \label{fig4}
\end{figure}

In a similar way, Figure \ref{fig4} displays the displacement snapshots for a 6-beam resonator.
The conclusions are equivalent to the previous case; when $\lambda \leq$ 4$h$, the biharmonic equation is not a good model representing the flexural waves propagating in a plate.
From these numerical experiments, we can conclude that the developed theory works well for resonators with arbitrary number of beams provided that the plate is thin enough to meet Kirchhoff's assumptions.
\subsection{Scattering properties of $N$-beam resonators}
Compared with an empty hole, the resonant structures under study are expected to show more complex scattering features.
These particular features are thoroughly illustrated in this section via the scattering cross-section introduced in Eq. (\ref{eqcs}), which is calculated by considering a plane wave propagating along the positive $x$-axis, $W=e^{ik_px}$, interacting with the $N$-beam resonator.
The cases of 1-beam and 2-beam resonators are studied with some details in what follows.
\begin{figure}[ht]
    \centering
    \includegraphics[scale=1]{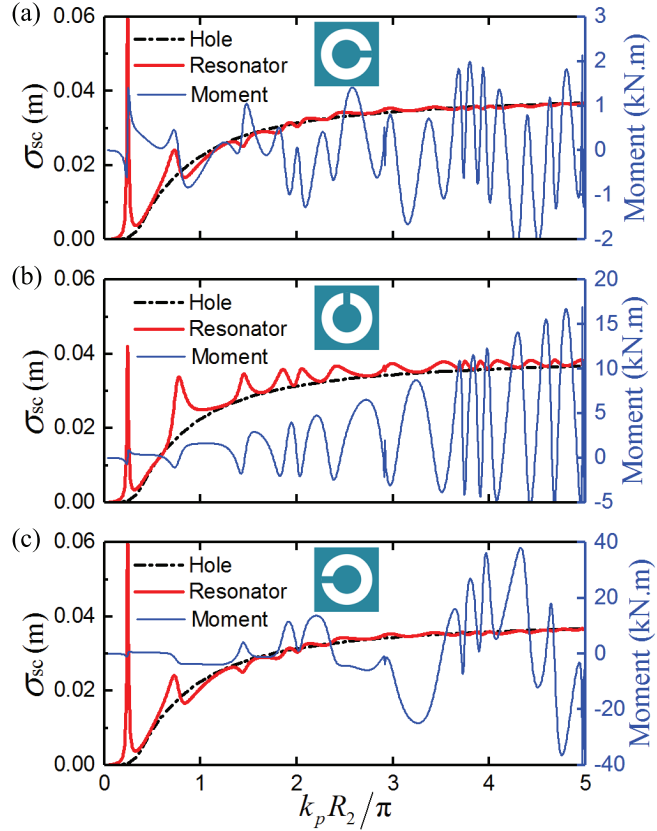}
	  \caption{(color on line) Frequency dependence of the scattering cross-section (in meters) of a single beam resonator, which is placed along the positive $x$-axis (a), perpendicularly to it (b) and opposite to it (c).
		The frequencies are given in normalized units, $k_{p}R_2/\pi$.
		Results are calculated for an empty hole with the size of the outer radius $R_2$ are given for comparison purposes.
		The blue thin lines represent the bending moments at the anchor points connected to the background plate.
		The insets show the alignment of the beam with respect to the incident wave, which propagates along the positive direction of the $x$-axis.
}
    \label{fig5}
\end{figure}
\par
First, the results for the single beam resonator are depicted in Fig. \ref{fig5}, which shows the dependence of $\sigma_{sc}$ on the normalized frequency $k_{p}R_2/\pi$.
 Results in  Fig. \ref{fig5}(a) correspond to the beam orientated along the positive direction of the $x$-axis [i.e., $\theta_1=0$ in Fig. \ref{fig1}].
Correspondingly, Figs. \ref{fig5}(b) and \ref{fig5}(c) represent the orientations $\theta_1=\pi$ /2 and $\theta_1=\pi$, respectively.
For comparison purpose, the dashed lines in both figures represent the results obtained for an empty hole with the same diameter.
It is observed that the curves for the three orientations are nearly overlapping with the dashed lines except some sharp peaks appearing at low frequencies.
Particularly, the first two peaks in all the figures appear at the same frequencies and are much stronger than the others.
In order to understand the physical origin of these peaks, Fig. \ref{fig5} also contains the bending moment $M_{x}^{1}(R_2)$ calculated at the anchor point $\Psi_{1,2}$ (continuous blue lines).
Clearly, the peaks of $\sigma_{sc}$ appear when the value of the bending moment experiences a strong variation, indicating the excitation of a resonant mode in the inner plate.
In other words, the large values of $\sigma_{sc}$ at the two frequencies are directly caused by the large (out-of-plane) displacements of the inner plate, their characteristics are described below.
\par
\begin{figure}[ht]
  \centering
  \includegraphics[scale=1]{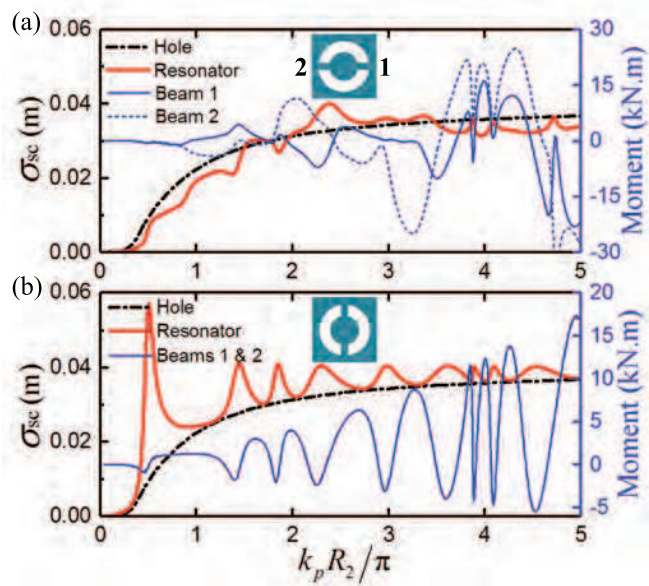}
	\caption{(color on line) Frequency dependence of $\sigma_{sc}$ (in meters) for the 2-beam resonator, which is placed with both beams orientated along the $x$-axis (a) and perpendicularly to it (b).
	The frequencies are given in normalized units, $k_{p}R_2/\pi$.
	The solid and dashed blue lines represent the bending moments at the two anchor points $\Psi_{1,2}$ and $\Psi_{2,2}$, respectively.
	They are degenerated in case (b).
	The results for an empty hole with the size of the outer radius $R_2$ are given for comparison purposes.
	The insets show the alignment of the beams with respect to the incident wave, which propagates along positive direction of the $x$-axis.}
  \label{fig6}
\end{figure}

Figure \ref{fig6} shows the corresponding results for a 2-beam resonator.
Since the scattering process is controlled by the combined action of beams, it is expected that an increasing number of beams will produce a rather complex behavior in $\sigma_{sc}$.
However, for the case of two beams the scattering properties of the resonator can be still understood in an intuitive manner.
For example, results in Fig. \ref{fig6}(a), corresponding to the beams aligned along the $x$-axis, show that the bending moments at the two anchor points ($\Psi_{1,2}$ and $\Psi_{2,2}$) are not excited in phase.
Consequently, the peaks of $\sigma_{sc}$ are weakened and their frequency positions cannot be related with the behavior of the bending moments.
The fact that the two beams are not oscillating in phase produces a frequency behavior on the cross-section closely similar to that of the empty hole.
\par
\begin{figure}[ht]
    \centering
    \includegraphics[scale=1]{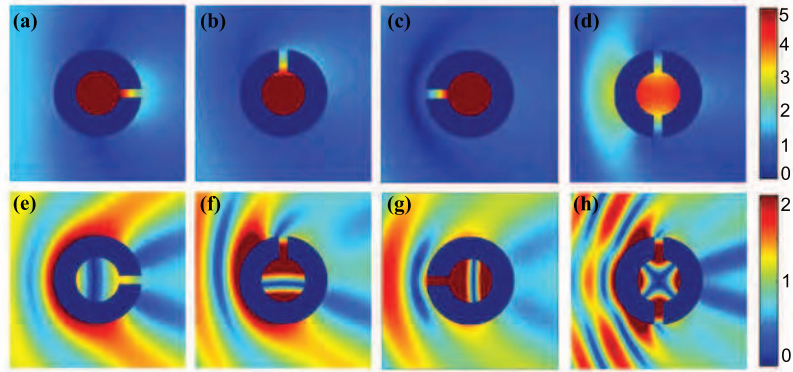}
	\caption{(color on line) Two-dimensional patterns of the (out-of-plate) displacement amplitude for the first peaks (upper panels) and second peaks (lower panels) in Figs. \ref{fig5} and \ref{fig6}.}
    \label{fig7}
\end{figure}

However, when the beams are perpendicular to the $x$-axis, the two anchor points are excited in phase with the wavefront, the bending moments are degenerated and $\sigma_{sc}$ shows many strong peaks.
Like for the two peaks analyzed in Figs. \ref{fig5}(a)-(b)-(c), the bending moments in Fig. \ref{fig6}(b) experience a rapid variation of their values for the two strongest peaks observed for $\sigma_{sc}$.
Again, this feature indicates the excitation of resonant modes inside the inner plate.
To prove this, Fig. \ref{fig7} shows two-dimensional patterns of the displacement amplitudes obtained at the first and second peaks in Figs. \ref{fig5} and \ref{fig6}.
Apparently, the patterns corresponding to the first peak [see Figs. \ref{fig7}(a), (b), (c) and (d)] demonstrate that the displacement of the inner plate is much larger than the background plate.
In fact, it is shown that the inner disk deforms as a whole in the out-of-plane direction just like the mass in a spring-mass resonator.
This movement can be considered as the fundamental resonance of the $N$-beam resonator.
\par
The frequency of the fundamental mode increases with the number of beams since the effective stiffness of our $N$-beam resonator increases with $N$.
Thus, from Figs. \ref{fig5} and \ref{fig6}(b) it is observed that the first peak experiences a blue shift when a beam is added to the resonator.
In reduced units, the frequency goes from 0.2404 to 0.5043, corresponding to 1407 Hz and 6175 Hz, respectively.
The value of the last frequency is in agreement with 6.1 kHz, the frequency of resonant band associated to the fundamental mode reported in Ref. \onlinecite{andreassen2015directional}, which studied square arrays of 2-beam resonators.
Quite differently, for the second peak, the 2D patterns in Figs. \ref{fig7}(e)-(f), (g) and (h) show that nodal lines appear inside the inner plate.
Like in the case of resonances of a circular diaphragm with clumped boundary conditions tabulated by Leissa, \cite{leissa1969vibration} the resonances inside our $N$-beam resonator can be also classified according to the number of nodal lines.
Thus, the single nodal line observed in the inner plate for Figs. \ref{fig7}(e), (f) and (g) indicates the excitation of the same resonance.
The two nodal lines shown in Fig. \ref{fig7}(h) denote the excitation of a resonant mode with higher order.
\par
The excitation of low frequency resonances is the origin of many interesting phenomena in flexural wave propagation.
For example, the single beam resonator studied in this work has been employed to propose elastic metamaterials with dispersive effective density.\citep{farhat2014biharmonic}
It is worth mentioning that the negative density is closely related with the rapid variation of bending moment at the resonance frequency.
\begin{figure}[ht]
    \centering
    \includegraphics[scale=1]{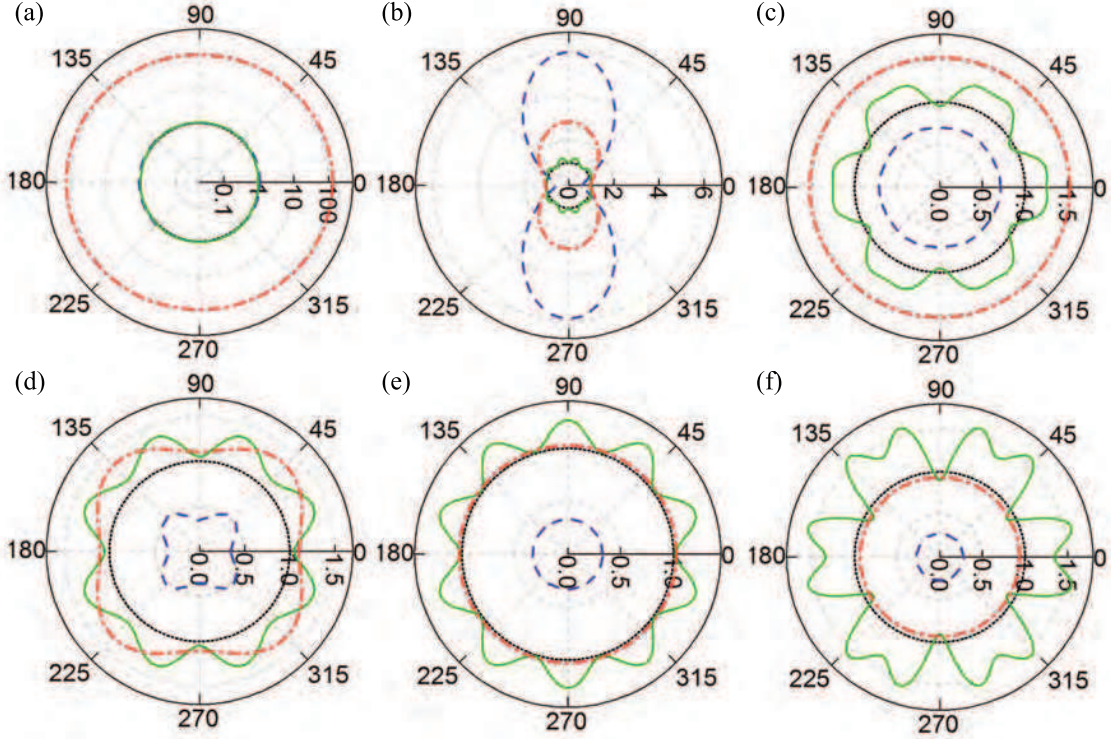}
		\caption{Polar plots depicting the angular variation of the normalized scattering cross-section, $\sigma_{sc}/\sigma_{sc}^{hole}$, from a $N$-beam resonator with 1- (a), 2- (b), 3- (c), 4- (d), 5- (e) and 6- (f) beams, respectively.
		Results are shown for three different frequencies, corresponding to wavelengths $\lambda =$ 8.3$R_2$ (red dot-dashed lines), 4$R_2$ (blue dashed lines) and $R_2$ (green lines).
	The circle with unity radius (black dotted lines) represents $\sigma_{sc}^{hole}$, used as the reference.}
    \label{fig8}
\end{figure}
\par
In contrast with a free circular hole, the scatterers under study don't possess radial symmetry, they are inherently anisotropic and have resonant modes at wavelengths larger than their geometrical dimensions (low frequencies).
Therefore, it is expected a frequency dependence of their directional scattering properties, which can be applied to propose devices for wave directionality and to design elastic metamaterials with low-frequency bandgaps. \cite{andreassen2015directional}
\par
The scattering properties of $N$-beam resonators, containing up to six beams, are reported in Figure \ref{fig8}, which shows the angular dependence of the normalized scattering cross-section, $\sigma_{sc}/\sigma_{sc}^{hole}$, with $\sigma_{sc}^{hole}$ being the cross-section of the empty hole (black dotted lines) with the same size.
For the sake of their comparison, calculations have been performed for three different frequencies: $k_{p}R_2=0.24\pi$ (red dashed-dotted lines), $k_{p}R_2=0.5\pi$ (blue dashed lines) and $k_{p}R_2=2\pi$ (green lines).
The first two frequencies correspond to the fundamental modes shown in Figs. \ref{fig5} and \ref{fig6}, with wavelengths $\lambda =$ 8.3$R_2$ and 4$R_2$, respectively.
\par
For the largest wavelength studied (red dot-dashed lines), Fig. \ref{fig8}(a) shows that $\sigma_{sc}$ is extremely large in comparison with $\sigma_{sc}^{hole}$ since it corresponds to the frequency of the fundamental mode of the 1-beam resonator.
In addition, it is observed that $\sigma_{sc}$ is almost isotropic since the mode can be excited independently of the direction of the incident wave as it is shown in Fig. \ref{fig5}.
However, for the 2-beam and 4-beam resonators [see Figs. \ref{fig8}(b) and \ref{fig8}(c), respectively], $\sigma_{sc}$ has a strong anisotropic character which is attributed to the fact that the beams are excited in couples, as it is shown in Fig. \ref{fig6} for the 2-beam resonator.
In other words, when the beams are symmetrically placed with respect to the incident direction (the $x$-axis), strong scattering occurs owing to the in phase vibration for couples of the beams.
For the 4-beam resonator this happens at four orientations; at 45$^o$ and its multiples.
Moreover, for these two resonators the cross-sections are larger than that of the empty hole since the frequency belongs to the peak of the fundamental mode.
For the other resonators, the beams are oscillating out-of-phase, producing an isotropic angular pattern with values of $\sigma_{sc}$ being larger or lower than $\sigma_{sc}^{hole}$.
\par
For intermediate wavelengths (blue dashed lines), the scattering by the 2- and 4-beam resonators keeps their anisotropic character for the reason explained above.
But now the $\sigma_{sc}$ value for the 2-beam case is about six times larger than $\sigma_{sc}^{hole}$ because the frequency exactly corresponds to its fundamental mode [see Fig. \ref{fig6}(b)].
For the 1-beam resonator case, $\sigma_{sc}\approx\sigma_{sc}^{hole}$ and is isotropic since there is no peak at this frequency (see Fig. \ref{fig5}).
The other resonators exhibit an isotropic behavior and have $\sigma_{sc}$ values always smaller than $\sigma_{sc}^{hole}$, decreasing with the number of beams.
This behavior can be understood if one considers that more beams mean more propagating channels for the incident flexural waves.
\par
Finally, for wavelengths comparable with the resonator size and below, like $\lambda=R_2$ (green lines), $\sigma_{sc}$ is strongly anisotropic for all the resonators except for the case of 1-beam, which has $\sigma_{sc}$ values practically independent of the beam orientation [see Fig. \ref{fig5}].
Thus, for high frequencies the inherent anisotropic of the scattering object is reflected in its angular dependence of $\sigma_{sc}$.
\par
From the results above, special discussion deserves the case for the 2-beam resonator [see Fig. \ref{fig8}(b)] that, for low frequencies, displays a characteristic "figure of eight" scattering pattern. This shape, indicating the angular dependence of the scattering with respect to the orientation of the exciting wave, can be employed in designing directional elastic metamaterials consisting of a perforated plate with a periodic distributions of them, as has been proven in Ref. \onlinecite{andreassen2015directional}.
\section{Summary}
In summary, the scattering of flexural waves from a hole containing a $N$-beam resonator has been solved analytically by using the impedance matrix method for flexural waves.
The model has been developed within the framework of the Kirchhoff-Love theory applied to thin plates.
 The expression for the transfer matrix (T-matrix) has been given as a function of the impedance matrices for the incident ($\boldsymbol{Z}^{inc}$), scattered ($\boldsymbol{Z}^{scat}$) and total ($\boldsymbol{Z}^{tot}$) waves.
For frequencies low enough, numerical results indicate that the transverse (out-of-plane) displacements are in agreement with 3D finite element simulations as well as with a 2D model formulated using Mindlin plate elements.\onlinecite{andreassen2015directional} It has been also determined the truncation of the series expression when dealing with impinging waves having high frequency.
Since the resonators are inherently anisotropic, they exhibit complex scattering characteristics, like directionality and resonances at low frequencies, which are revealed by studying the scattering cross-section.
Of particular interest is the case of 2-beam resonator, which exhibits a characteristics "figure of eight" scattering pattern at low frequencies.
This theoretical work opens the door to tackle a variety of problems within the field of elastic metamaterials based on perforated plates containing periodic distributions of $N$-beam resonators.
The development of a homogenization theory for this type of metamaterials and the demonstration of negative refraction based on them are the goals of further works to be published elsewhere.
\par
\section*{Acknowledgements}
This work was supported by the Ministerio de Economía y Competitividad of the Spanish government and the European Union Fondo Europeo de Desarrollo
Regional (FEDER) through Project No. TEC2014-53088-C3-1-R, and the National Science Foundation of China under Grant No. 11432004. Penglin Gao acknowledges a scholarship with No.201606120070 provided by China Scholarship Council.
\appendix
\numberwithin{equation}{section}
\numberwithin{figure}{section}
\section{Impedance matrices in region II: $\textbf{Z}^{inc}$ and $\textbf{Z}^{scat}$} \label{plate}
The incident and scattered impedance matrices in the infinite plate are defined by
\begin{equation}
\label{inc_matrix}
\begin{bmatrix} M_r^{\uppercase\expandafter{\romannumeral2}} \\ V_r^{\uppercase\expandafter{\romannumeral2}}\end{bmatrix}_q^{(inc)}=
\sum_{s=-\infty}^{\infty} -\boldsymbol{Z}_{qs}^{(inc)}
\begin{bmatrix} W^{\uppercase\expandafter{\romannumeral2}} \\ W_{,r}^{\uppercase\expandafter{\romannumeral2}}\end{bmatrix}_s^{(inc)},
\end{equation}
\begin{equation}
\label{sca_matrix}
\begin{bmatrix} M_r^{\uppercase\expandafter{\romannumeral2}} \\ V_r^{\uppercase\expandafter{\romannumeral2}}\end{bmatrix}_q^{(scat)}=
\sum_{s=-\infty}^{\infty} \boldsymbol{Z}_{qs}^{(scat)}
\begin{bmatrix} W^{\uppercase\expandafter{\romannumeral2}} \\ W_{,r}^{\uppercase\expandafter{\romannumeral2}}\end{bmatrix}_s^{(scat)},
\end{equation}
Based on these definitions, the two matrices can be expressed as
\begin{equation}
    \boldsymbol{Z}_{qq}^{inc}=-\boldsymbol{N}_{qq}^{JI}(k_p,R_2)\left [ \boldsymbol{M}_{qq}^{JI}(k_p,R_2) \right]^{-1},
\end{equation}
\begin{equation}
    \boldsymbol{Z}_{qq}^{scat}=\boldsymbol{N}_{qq}^{HK}(k_p,R_2)\left [ \boldsymbol{M}_{qq}^{HK}(k_p,R_2) \right]^{-1},
\end{equation}
where the additional matrices appearing in the above equations are defined as
\begin{equation}
\boldsymbol{M}_{qq}^{\Upsilon\Phi}(k_i,r)=
\begin{bmatrix}
    \Upsilon_{q}(k_i r) & \Phi_{q}(k_i r)\\
    k_i\Upsilon'_{q}(k_i r) & k_i\Phi'_{q}(k_i r),
\end{bmatrix}
\end{equation}
\begin{equation}
\boldsymbol{N}_{qq}^{\Upsilon\Phi}(k_i,r)=
\begin{bmatrix}
    S_{q}^{\Upsilon}(k_i r) & S^{\Phi}_{q}(k_i r)\\
    T_{q}^{\Upsilon}(k_i r) & T^{\Phi}_{q}(k_i r).
\end{bmatrix}
\end{equation}
\noindent where $S$ and $T$ represent complex expressions involving combinations of Bessel functions. Their explicit expressions are available in Ref. \onlinecite{climente2015scattering}.
\section{Impedance matrix in region II: \textbf{Z}$^{tot}$}\label{plate2}
The impedance matrix for the total wave propagating in the infinite plate is the key ingredient in the impedance method.
 \begin{equation}
    \label{definition_Zint}
    \left( \begin{bmatrix} M_r^{\uppercase\expandafter{\romannumeral2}} \\ V_r^{\uppercase\expandafter{\romannumeral2}}\end{bmatrix}_q^{inc}+
    \begin{bmatrix} M_r^{\uppercase\expandafter{\romannumeral2}} \\ V_r^{\uppercase\expandafter{\romannumeral2}}\end{bmatrix}_q^{scat} \right)=
    \sum_{s=-\infty}^{\infty} -\boldsymbol{Z}_{qs}^{tot}
    \left( \begin{bmatrix} W^{\uppercase\expandafter{\romannumeral2}} \\ W_{,r}^{\uppercase\expandafter{\romannumeral2}}\end{bmatrix}_s^{inc} +
    \begin{bmatrix} W^{\uppercase\expandafter{\romannumeral2}} \\ W_{,r}^{\uppercase\expandafter{\romannumeral2}}\end{bmatrix}_s^{scat} \right).
\end{equation}
This matrix is the key ingredient in the impedance method developed in Sec. II.
It is obtained by applying the modal expansion method in cylindrical coordinates to the boundary conditions.
\par
At the inner boundary $r=R_1$, for every $\theta$ we can define
\begin{equation}
    \label{innerplateW}
    \begin{bmatrix} W^{\uppercase\expandafter{\romannumeral1}} \\ W_{,r}^{\uppercase\expandafter{\romannumeral1}}\end{bmatrix}_{q}=
    \boldsymbol{M}^{JI}_{qq}(k_p,R_1)\begin{bmatrix} C_q^{(J)} \\ C_q^{(I)} \end{bmatrix},
\end{equation}
\noindent with $W_{,r}^{\uppercase\expandafter{\romannumeral1}}\equiv\partial W^{\uppercase\expandafter{\romannumeral1}}/\partial r$.

\begin{equation}
    \label{innerplateM}
    \begin{bmatrix} M_r^{\uppercase\expandafter{\romannumeral1}} \\ V_{r}^{\uppercase\expandafter{\romannumeral1}}\end{bmatrix}_{q}=
    \boldsymbol{N}^{JI}_{qq}(k_p,R_1)\begin{bmatrix} C_q^{(J)} \\ C_q^{(I)} \end{bmatrix}.
\end{equation}
Finally, the relation between Eqs. (\ref{innerplateW}) and (\ref{innerplateM}) is
\begin{equation}
    \label{Yqq}
    \begin{bmatrix} M_r^{\uppercase\expandafter{\romannumeral1}} \\ V_{r}^{\uppercase\expandafter{\romannumeral1}}\end{bmatrix}_{q}=
    (\boldsymbol{Y}^{\uppercase\expandafter{\romannumeral1}}_{qq})^{-1}
    \begin{bmatrix} W^{\uppercase\expandafter{\romannumeral1}} \\ W_{,r}^{\uppercase\expandafter{\romannumeral1}}\end{bmatrix}_{q},
\end{equation}
with
\begin{equation}
    \boldsymbol{Y}^{\uppercase\expandafter{\romannumeral1}}_{qq}=\boldsymbol{M}^{JI}_{qq}(k_p,R_1) \left[\boldsymbol{N}_{qq}^{JI}(k_p,R_1) \right]^{-1}.
\end{equation}
From the boundary conditions in Eqs. (\ref{bdc}) and (\ref{bdd}) (left column), we obtain that
      \begin{equation}
      \label{a}
      M_r^{\uppercase\expandafter{\romannumeral1}}(R_1,\theta)=\frac{1}{R_1}\sum_{n=1}^{N} M_x^{n}(R_1)\delta(\theta-\theta_{n}),
      \end{equation}
     \begin{equation}
     \label{b}
     V_r^{\uppercase\expandafter{\romannumeral1}}(R_1,\theta)=\frac{1}{R_1}\sum_{n=1}^{N} Q_x^{n}(R_1)\delta(\theta-\theta_{n}).
     \end{equation}
Expanding the Dirac function in azimuthal orders $\delta(\theta-\theta_n)=(1/2\pi)\sum_q e^{iq(\theta-\theta_n)}$, after grouping we obtain
\begin{equation}
    \label{Yqq2}
    \begin{bmatrix} M_r^{\uppercase\expandafter{\romannumeral1}}(R_1,\theta) \\ V_r^{\uppercase\expandafter{\romannumeral1}}(R_1,\theta) \end{bmatrix}=
    \sum_{q=-\infty}^{\infty}
    \begin{bmatrix} M_r^{\uppercase\expandafter{\romannumeral1}} \\ V_r^{\uppercase\expandafter{\romannumeral1}} \end{bmatrix}_{q} e^{iq\theta}
    =\sum_{q=-\infty}^{\infty} \left ( \sum_{n=1}^{N} \frac{e^{-iq\theta_n}}{2\pi R_1} \begin{bmatrix} M_x^{n}(R_1) \\ Q_x^{n}(R_1) \end{bmatrix} \right) e^{iq\theta}.
\end{equation}
Considering only the $q$-term of Eqs. (\ref{Yqq}) and (\ref{Yqq2}), we have
\begin{equation}
    \begin{bmatrix} W^{\uppercase\expandafter{\romannumeral1}} \\ W_{,r}^{\uppercase\expandafter{\romannumeral1}}\end{bmatrix}_{q}=
    \boldsymbol{Y}_{qq}^{\uppercase\expandafter{\romannumeral1}} \left ( \sum_{n=1}^{N} \frac{e^{-iq\theta_n}}{2\pi R_1} \begin{bmatrix} M_x^{n}(R_1) \\ Q_x^{n}(R_1) \end{bmatrix} \right).
\end{equation}

From the boundary conditions in Eqs. (\ref{bda}) and (\ref{bdb}) (left column), the vertical displacement and slope of the $m$-th beam can be expressed as function of bending moment and shear force
\begin{equation}
    \label{MW}
    \begin{bmatrix} V_x^{m}(R_1) \\ V_{,x}^{m}(R_1) \end{bmatrix}=
    \sum_{q=-\infty}^{\infty}\sum_{n=1}^{N}\boldsymbol{Y}_{qq}^{\uppercase\expandafter{\romannumeral1}}
    \frac{e^{iq(\theta_m-\theta_n)}}{2\pi R_1}
    \begin{bmatrix} M_x^{n}(R_1) \\ Q_{x}^{n}(R_1) \end{bmatrix}.
\end{equation}
Notice that $\boldsymbol{K}_{ij}$ are not dependent on $n$. Rewritten Eq. (\ref{stiffness}) to fit the need of Eq. (\ref{MW}) as
\begin{equation}
    \begin{bmatrix} V_x^{n}(R_1) \\ V_{,x}^{n}(R_1)\end{bmatrix}=
    \boldsymbol{K}^{-1}_{21}\begin{bmatrix} M_x^{n}(R_2) \\ Q_{x}^{n}(R_2)\end{bmatrix}-
    \boldsymbol{K}^{-1}_{21}\boldsymbol{K}_{22}\begin{bmatrix} V_x^{n}(R_2) \\ V_{,x}^{n}(R_2)\end{bmatrix},
\end{equation}
\begin{equation}
    \begin{bmatrix} M_x^{n}(R_1) \\ Q_{x}^{n}(R_1)\end{bmatrix}=
    \boldsymbol{K}_{11}\boldsymbol{K}^{-1}_{21}\begin{bmatrix} M_x^{n}(R_2) \\ Q_{x}^{n}(R_2)\end{bmatrix}+
    (\boldsymbol{K}_{12}-\boldsymbol{K}_{11}\boldsymbol{K}^{-1}_{21}\boldsymbol{K}_{22})\begin{bmatrix} V_x^{n}(R_2) \\ V_{,x}^{n}(R_2)\end{bmatrix},
\end{equation}
and casting them into Eq. (\ref{MW}), we obtain the following equation after some manipulations
\begin{equation}
    \label{linear_set}
    \sum_{n=1}^{N}\boldsymbol{L}_{mn}\begin{bmatrix} M_x^{n}(R_2) \\ Q_{x}^{n}(R_2)\end{bmatrix}=
    \sum_{n=1}^{N}\boldsymbol{R}_{mn}\begin{bmatrix} V_x^{n}(R_2) \\ V_{,x}^{n}(R_2)\end{bmatrix}.
\end{equation}
The following notations have been used to simplify the equation
\begin{subequations}
\begin{equation}
    \boldsymbol{L}_{mn}=\boldsymbol{\delta}_{mn}\boldsymbol{K}_{21}^{-1}-
    \boldsymbol{Y}_{mn}^{\uppercase\expandafter{\romannumeral1}}\boldsymbol{K}_{11}\boldsymbol{K}_{21}^{-1},
\end{equation}
\begin{equation}
    \boldsymbol{R}_{mn}=\boldsymbol{\delta}_{mn}\boldsymbol{K}_{21}^{-1}\boldsymbol{K}_{22}+
    \boldsymbol{Y}_{mn}^{\uppercase\expandafter{\romannumeral1}}(\boldsymbol{K}_{12}-
    \boldsymbol{K}_{11}\boldsymbol{K}_{21}^{-1}\boldsymbol{K}_{22}),
\end{equation}
\begin{equation}
    \boldsymbol{Y}_{mn}^{\uppercase\expandafter{\romannumeral1}}=\frac{1}{2\pi R_1}
    \sum_{q=-\infty}^{\infty}\boldsymbol{Y}_{qq}^{\uppercase\expandafter{\romannumeral1}}e^{iq(\theta_m-\theta_n)},
\end{equation}
\end{subequations}
where the sub-matrices $\boldsymbol{L}_{mn}$ and $\boldsymbol{R}_{mn}$ are the elements of block matrices $\boldsymbol{L}$ and $\boldsymbol{R}$, respectively. $\boldsymbol{\delta}_{mn}$ is the delta function in matrix form and indices $m,n\in[1,N]$. Solving the linear system of equations in Eq. (\ref{linear_set}), we obtain
\begin{equation}
    \label{innerimpedance}
    \begin{bmatrix} M_x^{m}(R_2) \\ Q_{x}^{m}(R_2)\end{bmatrix}=\sum_{n=1}^{N}\boldsymbol{Z}_{mn}
    \begin{bmatrix} V_x^{n}(R_2) \\ V_{,x}^{n}(R_2)\end{bmatrix},
\end{equation}
where $\boldsymbol{Z}_{mn}$ is the element of auxiliary impedance matrix $\boldsymbol{Z}=\boldsymbol{L}^{-1}\boldsymbol{R}$.

As have been done in Eqs. (\ref{a})-(\ref{Yqq2}), the boundary conditions in Eqs. (\ref{bdc}) and (\ref{bdd}) (right column) are expanded in azimuthal orders
\begin{equation}
    \label{outer}
    \begin{bmatrix} M_r^{\uppercase\expandafter{\romannumeral2}}(R_2,\theta) \\ V_r^{\uppercase\expandafter{\romannumeral2}}(R_2,\theta) \end{bmatrix}^{tot}=
    \sum_{q=-\infty}^{\infty}
    \begin{bmatrix} M_r^{\uppercase\expandafter{\romannumeral2}} \\ V_r^{\uppercase\expandafter{\romannumeral2}} \end{bmatrix}_{q}^{tot} e^{iq\theta}
    =\sum_{q=-\infty}^{\infty} \left ( \sum_{m=1}^{N} \frac{e^{-iq\theta_m}}{2\pi R_2} \begin{bmatrix} M_x^{m}(R_2) \\ Q_x^{m}(R_2) \end{bmatrix} \right) e^{iq\theta}.
\end{equation}
Taking into account the $q$-term only and we obtain the following relationship after substituting Eq. (\ref{innerimpedance}) into it
\begin{equation}
    \begin{bmatrix} M_r^{\uppercase\expandafter{\romannumeral2}} \\ V_r^{\uppercase\expandafter{\romannumeral2}} \end{bmatrix}_{q}^{tot}=\sum_{m=1}^{N}\sum_{n=1}^{N} \frac{e^{-iq\theta_m}}{2\pi R_2}\boldsymbol{Z}_{mn} \begin{bmatrix} V_x^{n}(R_2) \\ V_{,x}^{n}(R_2) \end{bmatrix}.
\end{equation}
Considering the boundary conditions in Eqs. (\ref{bda}) and (\ref{bdb}) (right column)
\begin{equation}
    \begin{bmatrix} V_x^{n}(R_2) \\ V_{,x}^{n}(R_2) \end{bmatrix}=\begin{bmatrix} W^{\uppercase\expandafter{\romannumeral2}}(R_2,\theta_n) \\ W_{,r}^{\uppercase\expandafter{\romannumeral2}}(R_2,\theta_n) \end{bmatrix}^{tot}=
    \sum_{s=-\infty}^{\infty} \begin{bmatrix} W^{\uppercase\expandafter{\romannumeral2}} \\ W_{,r}^{\uppercase\expandafter{\romannumeral2}} \end{bmatrix}_{s}^{tot}e^{is\theta_n},
\end{equation}
then
\begin{equation}
    \begin{bmatrix} M_r^{\uppercase\expandafter{\romannumeral2}} \\ V_r^{\uppercase\expandafter{\romannumeral2}} \end{bmatrix}_{q}^{tot}=\sum_{s=-\infty}^{\infty}\sum_{m=1}^{N}\sum_{n=1}^{N} \frac{e^{i(s\theta_n-q\theta_m)}}{2\pi R_2}\boldsymbol{Z}_{mn}
    \begin{bmatrix} W^{\uppercase\expandafter{\romannumeral2}} \\ W_{,r}^{\uppercase\expandafter{\romannumeral2}} \end{bmatrix}_{s}^{tot}.
\end{equation}
According to the definition in Eq. (\ref{definition_Zint}), the $2\times2$ blocks of the impedance matrix for the total wave are expressed as
\begin{equation}
    \boldsymbol{Z}_{qs}^{tot}=-\sum_{m=1}^{N}\sum_{n=1}^{N} \frac{e^{i(s\theta_n-q\theta_m)}}{2\pi R_2}\boldsymbol{Z}_{mn}.
\end{equation}

\section{Beam stiffness matrix}
The background plate and inner circular plate are connected by the $N$-beams, which behave like springs.
This appendix describes how to connect the bending moment and shear force to the displacement and slope at the anchor points by the stiffness matrix.
\par
Let us consider an arbitrary beam. After differentiating Eq. (\ref{eq_beam}) multiple times with respect to coordinate $x$, we obtain
\begin{equation}
	\begin{bmatrix}V   (x)\\ V'  (x)\\ V'' (x)\\ V'''(x) \end{bmatrix}	=
	\begin{bmatrix}
		(ik_b)^{0}e^{ik_b x} & (-ik_b)^{0}e^{-ik_b x} & (k_b)^{0}e^{k_b x} & (-k_b)^{0}e^{-k_b x} \\
		(ik_b)^{1}e^{ik_b x} & (-ik_b)^{1}e^{-ik_b x} & (k_b)^{1}e^{k_b x} & (-k_b)^{1}e^{-k_b x} \\
		(ik_b)^{2}e^{ik_b x} & (-ik_b)^{2}e^{-ik_b x} & (k_b)^{2}e^{k_b x} & (-k_b)^{2}e^{-k_b x} \\
		(ik_b)^{3}e^{ik_b x} & (-ik_b)^{3}e^{-ik_b x} & (k_b)^{3}e^{k_b x} & (-k_b)^{3}e^{-k_b x}
	\end{bmatrix}
	\begin{bmatrix} D_{1}\\ D_{2}\\ D_{3}\\ D_{4} \end{bmatrix}.
\end{equation}
Evaluating this at the end points $x=R_1$ and $x=R_2$, after simplifying and ordering terms we have
\begin{equation}
	\begin{bmatrix}V   (R_1)\\ V'  (R_1) \\ V   (R_2)\\ V'  (R_2) \end{bmatrix}	=
	\begin{bmatrix}
		\quad	e^{ik_b R_1} & \quad	e^{-ik_b R_1}  & \quad e^{k_b R_1} & \quad e^{ -k_b R_1} \\
		ik_b	e^{ik_b R_1} & -ik_b	e^{ -ik_b R_1}  & k_b   e^{k_b R_1}	&	-k_b	e^{ -k_b R_1} \\
		\quad e^{ ik_b R_2} & \quad e^{-ik_b R_2}  & \quad e^{ k_b R_2} & \quad	e^{-k_b R_2} \\
		ik_b	e^{ ik_b R_2} & -ik_b	e^{-ik_b R_2}  & k_b   e^{ k_b R_2} & -k_b	e^{-k_b R_2}
	\end{bmatrix}
	\begin{bmatrix} D_{1}\\ D_{2}\\ D_{3}\\ D_{4} \end{bmatrix} =
					\boldsymbol{H}	\begin{bmatrix} D_{1}\\ D_{2}\\ D_{3}\\ D_{4} \end{bmatrix},
\end{equation}
\begin{equation}
	\begin{bmatrix}V'' (R_1)\\ V'''(R_1) \\ V'' (R_2)\\ V'''(R_2)\end{bmatrix}	= k_b^2
	\begin{bmatrix}
		\quad -e^{ik_b R_1} & \quad -e^{ -ik_b R_1}   & \quad e^{k_b R_1}  & \quad e^{ -k_b R_1} \\
		-ik_b	 e^{ik_b R_1} &  ik_b	e^{ -ik_b R_1}   & k_b	  e^{k_b R_1}  & -k_b	e^{ -k_b R_1} \\
		\quad -e^{ ik_b R_2} & \quad -e^{-ik_b R_2} 	& \quad e^{ k_b R_2} 	& \quad e^{-k_b R_2} \\
		-ik_b	 e^{ ik_b R_2} &  ik_b	e^{-ik_b R_2} 	& k_b	  e^{ k_b R_2} 	& -k_b	e^{-k_b R_2}
	\end{bmatrix}
	\begin{bmatrix} D_{1}\\ D_{2}\\ D_{3}\\ D_{4} \end{bmatrix}=
	k_b^2  	\boldsymbol{H} \boldsymbol{\alpha} \begin{bmatrix} D_{1}\\ D_{2}\\ D_{3}\\ D_{4} \end{bmatrix},
\end{equation}
where $\boldsymbol{\alpha}=$ diag($-1$, $-1$, $1$, $1$) is a diagonal matrix. Combining the previous equations and after some tedious algebra operations, we obtain the following relation:
\begin{equation}
    \label{stiffness}
	\begin{bmatrix}
		M_x(R_1) \\ Q_x(R_1) \\ M_x(R_2) \\ Q_x(R_2)
	\end{bmatrix} =
	-EIk_b^2 \boldsymbol{H}\boldsymbol{\alpha}\boldsymbol{H}^{-1}
	\begin{bmatrix}
		 V(R_1) \\ V'(R_1) \\ V(R_2) \\ V'(R_2)
	\end{bmatrix} =
    \boldsymbol{K}
    \begin{bmatrix}
		 V(R_1) \\ V'(R_1) \\ V(R_2) \\ V'(R_2)
	\end{bmatrix}
\end{equation}
\noindent where the stiffness matrix is
\begin{equation}
    \boldsymbol{K}=
    \begin{bmatrix}
        \boldsymbol{K_{11}} & \boldsymbol{K_{12}} \\
        \boldsymbol{K_{21}} & \boldsymbol{K_{22}} \\
    \end{bmatrix}
    =-\frac{EIk_{b}^{2}}{S}
    \begin{bmatrix}
    S_{10} & \frac{S_{11}}{k_{b}} & S_{20} & \frac{S_{21}}{k_{b}} \\
    k_{b}S_{12} & -S_{10} & k_{b}S_{22} & S_{20} \\
    S_{20} & -\frac{S_{21}}{k_{b}} & S_{10} & -\frac{S_{11}}{k_{b}}\\
    -k_{b}S_{22} & S_{20} & -k_{b}S_{12} & -S_{10}\\
    \end{bmatrix},
\end{equation}
\noindent with notations
\begin{align}
    S&=-4e^{(1+i)k_{b}(R_1+R_2)}+(e^{2ik_b R_1}+e^{2ik_b R_2})(e^{2k_b R_1}+e^{2k_b R_2}),\\
    S_{10}&=-i\left[ e^{2(1+i)k_b R_1}+e^{2(1+i)k_b R_2}-e^{2k_b(R_1+iR_2)}-e^{2k_b(R_2+iR_1)}\right],\\
    S_{11}&=(1+i)\left[ e^{2(1+i)k_b R_1}-e^{2(1+i)k_b R_2}-ie^{2k_b(R_1+iR_2)}+ie^{2k_b(R_2+iR_1)}\right],\\
    S_{12}&=(1-i)\left[ e^{2(1+i)k_b R_1}-e^{2(1+i)k_b R_2}+ie^{2k_b(R_1+iR_2)}-ie^{2k_b(R_2+iR_1)}\right],\\
    S_{20}&=4e^{(1+i)k_b(R_1+R_2)}\big\{ \cos\left[k_b(R_1-R_2)\right]-\cosh\left[k_b(R_1-R_2)\right]\big\},\\
    S_{21}&=4e^{(1+i)k_b(R_1+R_2)}\big\{ \sin\left[k_b(R_1-R_2)\right]-\sinh\left[k_b(R_1-R_2)\right]\big\},\\
    S_{22}&=-4e^{(1+i)k_b(R_1+R_2)}\big\{ \sin\left[k_b(R_1-R_2)\right]+\sinh\left[k_b(R_1-R_2)\right]\big\}.
\end{align}

\end{document}